\documentclass[aps,prb,twocolumn,showpacs,preprintnumbers,amsmath,amssymb,superscriptaddress]{revtex4}
\usepackage{mathptm}  
\usepackage{dcolumn}                    
\usepackage{bm}                        
\usepackage{graphicx}
\usepackage{times}
\begin{document}
\newcommand {\be}{\begin{equation}}
\newcommand {\ee}{\end{equation}}
\newcommand {\bea}{\begin{eqnarray}}
\newcommand {\eea}{\end{eqnarray}}
\newcommand {\nn}{\nonumber}
\newcommand {\bb}{\bibitem}
\newcommand{\et}{{\,\it et al,\,\,}}

\title{Proposed experiment: out-of-plane nodal lines in Sr$_{2}$RuO$_{4}$}

\author{David Parker}
\affiliation{U.S. Naval Research Laboratory  - Code 6390,  4555 Overlook Ave. SW, Washington DC 20375}

\date{\today}

\begin{abstract}

Since the original proposal of an unconventional chiral order 
parameter in the ruthenate perovskite superconductor
Sr$_{2}$RuO$_{4}$, much attention has been given 
to the possibility of out-of-plane nodal lines on the predominant 
$\gamma$ cylindrical Fermi surface given evidence 
for low-lying quasiparticle
excitations in this material.  Here I propose a tunneling spectroscopy 
experiment to determine whether such nodal lines in fact exist.

\end{abstract}
\maketitle

{\it Introduction. }  Superconductivity at approximately 1 K was discovered in Sr$_{2}$RuO$_{4}$
in 1994 by Maeno et al \cite{maeno}, and since that time has been a topic of 
strong interest, with substantial experimental and theoretical activity 
continuing fifteen years after its discovery.  This material was found following a lengthy search for high-temperature
materials structurally similar to the high-T$_{c}$ 
cuprates, but not containing Cu.  It was thought \cite{mackenzie} 
that a new family of high-temperature superconductors might be discovered
in this way, and while this has not happened, interest in this material remains high.

Almost immediately after its discovery, Rice and Sigrist \cite{sigrist} proposed 
that this material contained a two-dimensional p-wave order
parameter as an `electronic
analogue' to superfluid Helium.  A closely related compound, SrRuO$_3$, shows 
ferromagnetism, and so the argument was made in analogy to the 
ferromagnetically mediated pairing in He.

Complicating this simple picture, however, is substantial evidence for 
nodal excitations in this material.  The proposed chiral order parameter would
give rise to low-temperature exponentially activated behavior in the various
thermodynamic quantities (such as magnetic penetration depth and nuclear spin relaxation rate), but this is not what has been observed.  Bonalde et al \cite{bonalde} measured the London penetration depth in single crystals of Sr$_{2}$RuO$_{4}$
and found a T$^{2}$ dependence ,while Nishizaki et al \cite{nishizaki} found
T$^{2}$ specific heat behavior, evidence for a line-node state.  Power-law behavior was also observed
in nuclear spin relaxation rate (T$_{1}^{-1}$) measurements \cite{ishida_old} and ultrasonic attenuation
\cite{lupien}.  
In addition, Izawa et al \cite{izawa} 
measured the magnetothermal conductivity of single crystals of Sr$_{2}$RuO$_{4}$ and found that any nodal
lines could not be parallel to the c-axis, which immediately suggested 
$\Delta({\bf k})=\exp(i\phi)\cos(ck_{z})$, given the previous evidence
for nodal excitations.   The lack of anisotropy in ab-plane magnetothermal conductivity measurements \cite{tanatar} also suggests any nodal lines are parallel to the basal plane. Most recently, Ishida \cite{ishida} et al again conducted T$_{1}^{-1}$ measurements on a high-quality sample of Sr$_{2}$RuO$_{4}$
and found $T^{3}$ behavior, commonly taken as indicative of line nodes.   Given these measurements, there may well be nodes parallel to the basal plane on 
Sr$_{2}$RuO$_{4}$. 

In this paper I propose an experiment that could help determine whether these nodal lines exist.  The method is based on tunneling spectroscopy, which can be a strong probe of order parameter symmetry.  
The basis of the experiment is presented in Figure 1 (reprinted from \cite{parker_graph}), which depicts, in
momentum space, an Sr$_{2}$RuO$_{4}$ c-axis tunneling spectroscopy experiment, based upon a recent proposal \cite{parker_graph} by the author and P. Thalmeier for the use of graphite as a normal electrode in a superconducting-insulator-graphite tunneling experiment.  As the method is described in detail in that publication I only sketch the proposal here.The basic idea is that the use of a gate voltage applied to the semi-metal graphite changes the length of the electron-occupied graphite HKH Fermi surface ``cigar".  When the graphite is deployed in an appropriate c-axis orientation, as indicated, the conservation of the momentum parallel to the interface $k_{\parallel}$ means that different cigar lengths will sample different regions of the Sr$_{2}$RuO$_{4}$ Fermi surface.  If the superconducting order parameter has no k$_{z}$-dependence, the sole effect of the lengthening of the cigar will be an an increase in conduction channels and thereby merely an overall scale factor in the conductance.  However, if the order parameter has k$_{z}$ dependence, each 
point on the cigar will see a region of different $\Delta({\bf k})$, producing tunneling or Andreev density of states features at that energy.  The differential conductance, when properly normalized to the high-bias value, would thus vary with gate voltage.

In this paper I limit myself to a proposal for experimental
detection of an $\exp(i\phi)\cos(k_{z})$ order parameter (OP).  
Several other gap functions have been proposed, including the
d-wave order parameter $\exp(i\phi)\sin(k_{z})$ \cite{zutic}, as well
as various other d-wave and f-wave OP's. The method described herein
for the $\exp(i\phi)\cos(k_{z})$ will in general yield distinguishable results for any OP
with significant k$_{z}$ dependence, but due to space constraints I present explicit results
only for this OP.

\begin{figure}[h!]
\includegraphics[width=4cm]{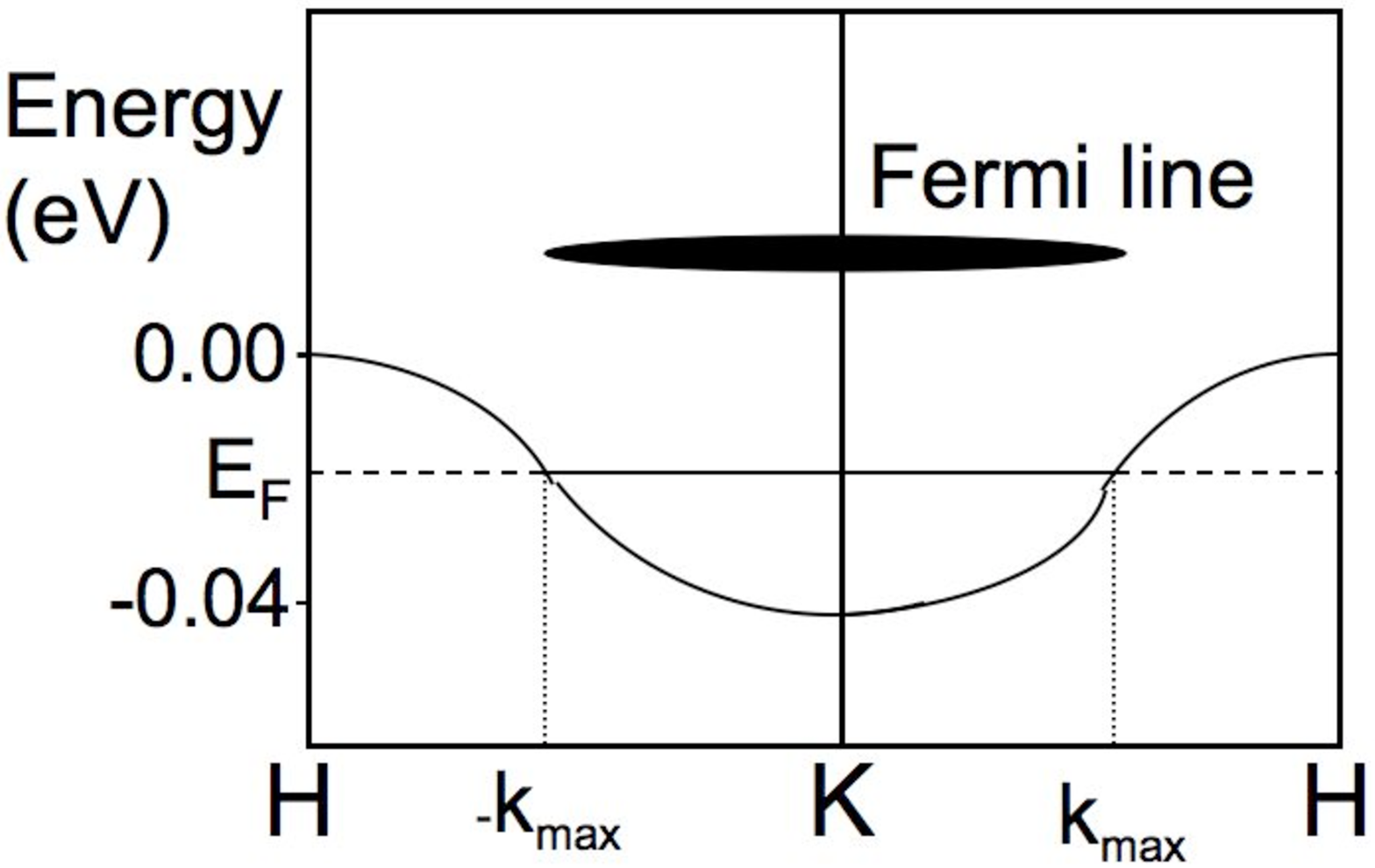}
\includegraphics[width=4cm]{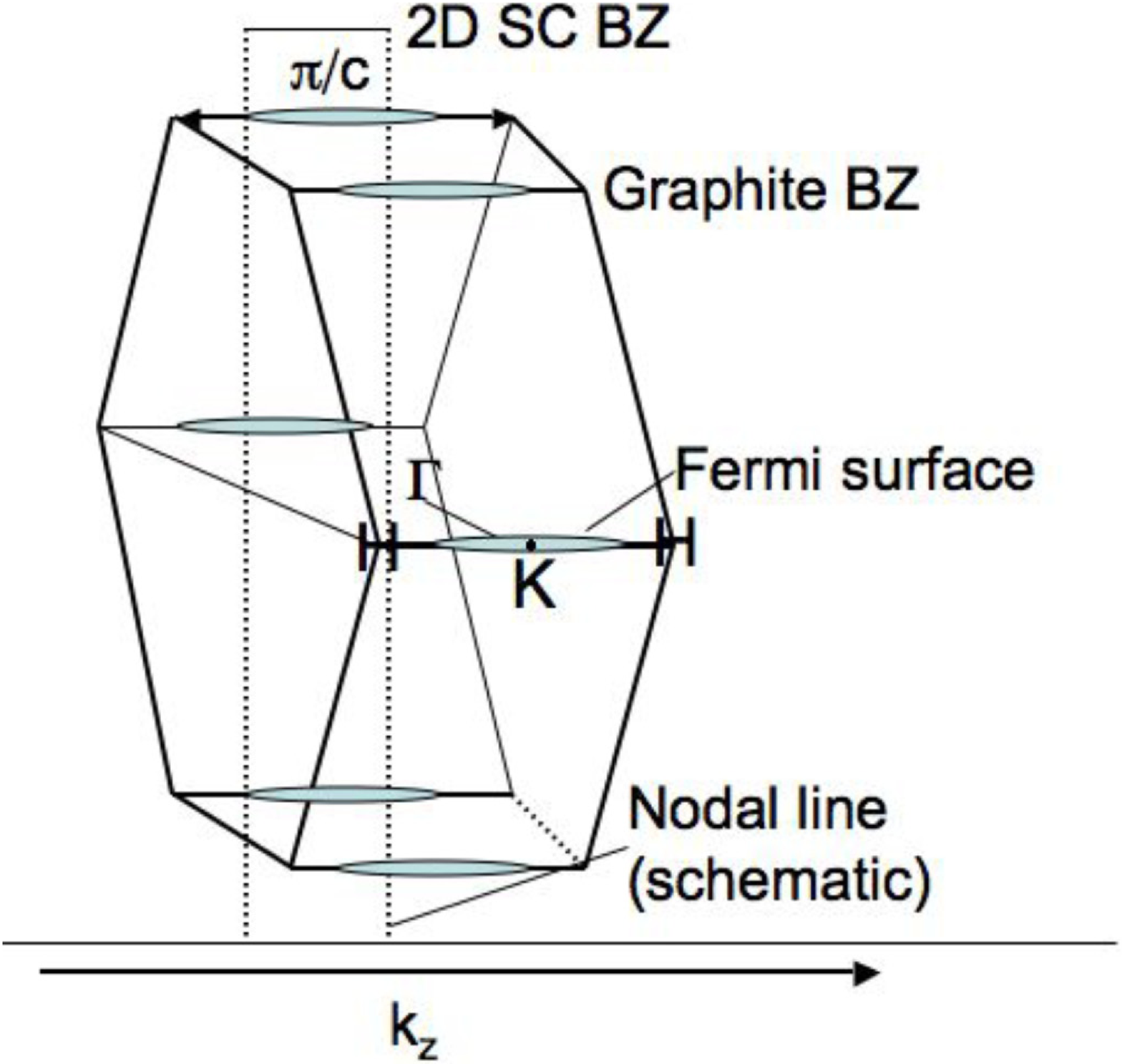}
\caption{A proposed tunneling spectroscopy experiment to help determine Sr$_{2}$RuO$_{4}$ order parameter symmetry.  Depending on the gate-voltage controlled length of the graphite Fermi HK line,
regions of different order parameter value are selected, leading to
different dI/dV behavior.}
\end{figure}

{\it Calculation.  } The calculation presented in this section follows the standard technique
applicable to Andreev and tunneling spectroscopy of anisotropic superconducting
order parameters \cite{blonder,tanaka,parker} and therefore I keep only the
most essential details of the calculation.
In general, the pair state of a superconductor is described by the
Bogoliubov-deGennes equations \cite{deGennes_book, blonder,tanaka,klapwijk_mar}:
\bea
i\hbar\frac{\partial f}{\partial t} &=& -\left[\frac{\hbar^{2}\nabla^{2}}{2m}+
\mu+V(x)\right]
f({\bf x},{\bf k},t) - \Delta({\bf x},{\bf k})g(x,t) \\
i\hbar\frac{\partial g}{\partial t} &=& \left[\frac{\hbar^{2}\nabla^{2}}{2m}+\mu+V(x)\right]g({\bf x},{\bf k},t) 
- \Delta({\bf x},{\bf k})f(x,t)
\eea
with f representing electron-like wavefunctions and g representing
hole-like wavefunctions, with solutions
\bea
f({\bf x},{\bf k},t)&=& u({\bf k})\exp(i\ ({\bf k}\cdot {\bf r} - Et)/\hbar) \\
g({\bf x},{\bf k},t)&=& v({\bf k})\exp(i\ ({\bf k}\cdot {\bf r} +Et)/\hbar)
\eea
with u and v are the BCS coherence factors \cite{BCSbook,
honerkamp}:
\bea
u({\bf k}) &=& \sqrt{\frac{1}{2}(1+\sqrt{E^{2}-|\Delta^{2}({\bf k})|}/E)} \\
v({\bf k}) &=& \exp(i\phi)\sqrt{\frac{1}{2}(1-\sqrt{E^{2}-|\Delta^{2}({\bf k})|}/E)}
\eea
Here $\phi$ is the phase of the gap $\Delta({\bf k})$.  Given an electron incident from the normal metal, two additional particles
result in the metal: an Andreev-reflected hole \cite{andreev}, and a normally
reflected electron, while in the superconductor an electron-like and hole-like
quasiparticle result.  Later in this work we allow for the effect
of quasiparticle scattering by letting the energy E have a finite
imaginary part $\Gamma$ \cite{dynes,plecenik,grajcar,parker}.  

Each of the particles above has a corresponding amplitude
(a, b,c and d, respectively) 
which is found by specifying the boundary conditions: 
continuity of the wavefunction across the boundary, and the following condition
applicable to $\delta$-function barrier potentials, as introduced in \cite{blonder}:
\bea 
\psi_{S}^{'}(0)-\psi_{N}^{'}(0)
 &=& \frac{2}{\hbar^{2}}H\psi(0)
\eea
with the barrier function potential H$\delta(x)$. Once a and b, the amplitudes for Andreev and normal reflection, have been
solved for the differential conductance dI/dV is calculated from the 
following:
\bea
dI/dV \propto \int dk_{\parallel} (1+|a(k_{\parallel},E)|^{2}-|b(k_{\parallel},E)|^{2})
\eea
The calculated results employ the boundary condition that k$_{\parallel}$,
the momentum parallel to the interface, 
is conserved.  This condition
allows for wavevector selection along the longitudinal Sr$_{2}$RuO$_{4}$
Fermi surface, and hence the acquisition of information about the order 
parameter value at this wavevector.
\begin{figure}[h!]
\includegraphics[width=6cm,angle=90]{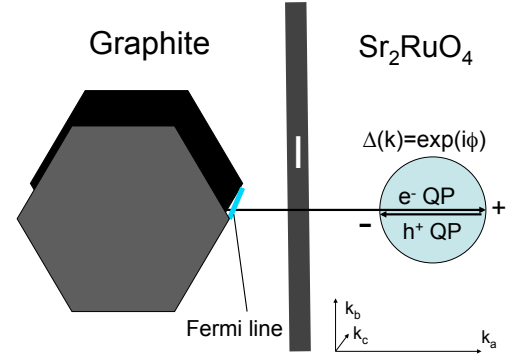}
\caption{The ab-plane layout of the proposed experiment in momentum-space.  Note that the real-space graphite axes are rotated 30$^o$ with respect to the momentum space axes.}
\end{figure}

We note that for such an ab-plane Sr$_{2}$RuO$_{4}$ tunneling experiment, the phase of the order
parameter is of crucial importance, as it determines the phase of the hole component wavefunction {\it v}.  Depicted in Figure 2 is an ab-plane diagram of the proposed experiment. The basic point is that assuming parallel momentum conservation, the electron-like quasiparticle will contact the Sr$_{2}$RuO$_{4}$ Fermi surface at k$_{b}=0$, k$_{a}=k_{F}$ sampling the order
parameter at this wavevector, while the backscattered hole will sample the wavevector directly opposite to this, which introduces a relative phase of -1 among the hole-like components of these quasiparticles, due to the order parameter sign change.  As shown by Tanaka \cite{tanaka}, such a sign change leads to a zero-bias conductance peak.  However, we will see that the character of this peak and surrounding features will depend substantially upon the gate voltage if the order parameter has $k_{z}$ dependence, leaving a characteristic signature of such a state.

\begin{figure}[h!]
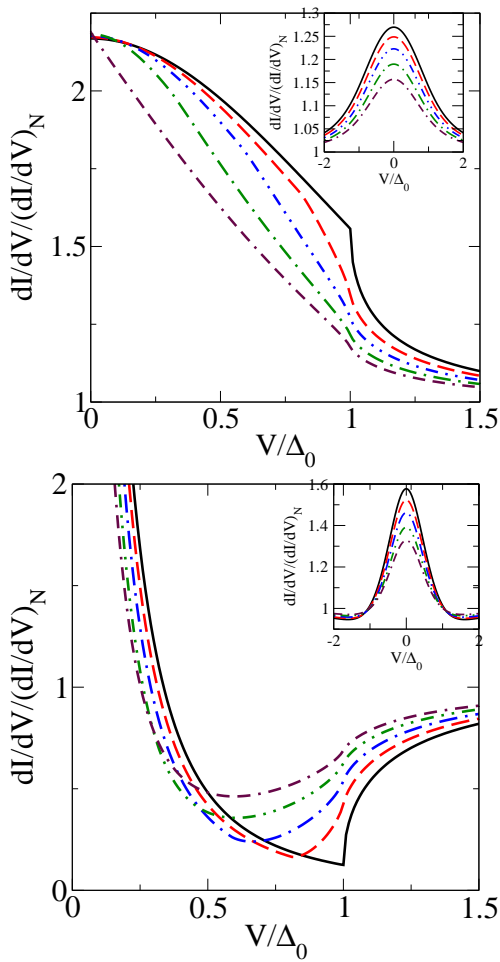

\includegraphics[width=6.5cm]{fig3a.eps}
\includegraphics[width=6.5cm]{fig3b.eps}
\caption{The results of dI/dV calculations in the Andreev limit (top, Z=0.3) and tunneling limit (bottom, Z=2), for graphite gated are shown. Graphite chemical potential changes $\delta\mu=-20meV$ (solid,black), $-18.8 meV$ (dashed,red), $-17.3 meV$ (dot-dashed, blue), $-15.3meV$ (double-dot-dashed, green) and $-12.7 meV$ (double-dash-dotted, maroon).  Insets: Dynes $\Gamma$ taken as 0.75 $\Delta_{0}$.}
\end{figure}

{\it Main Result.}  Depicted in Figure 3 is the main result of this paper.  The differential 
conductance dI/dV is shown in two limits, for several gate-induced graphite 
chemical potential changes for the barrier parameter Z: top, Z=0.3, corresponding
to the point contact regime, and bottom, Z=2, corresponding to the tunneling regime.  

Substantial effects of the gating are apparent in both plots.  In the top plot, the low-energy Andreev
signal evolves from a rounded hump containing a sharp feature at $V=\Delta_{0}$ towards a nearly linear behavior for $\delta\mu=-12.7 meV$, while in the bottom plot the depression in dI/dV that may sometimes occur adjacent to a ZBCP, most prominent for $\delta\mu=-20meV$ gradually fills in as the gate voltage increases and lower energy states are accessed, additionally narrowing the peak itself. 

The evolution of the curves with gate voltage is easily understood.  For the Andreev-limit plot, the width of the Andreev reflection (AR) signal narrows with increasing gate voltage because one is seeing an 
AR signal from a portion of Fermi surface with smaller $\Delta({\bf k})$ than the maximum gap, and the sharp feature present at $V=\Delta_{0}$ for $\delta\mu$=-20meV similarly becomes less prominent.
Analogously, for the tunneling limit the width of the ZBCP decreases with increasing gate voltage, and the feature at $V=\Delta_{0}$ is washed out by the summing of ZBCP curves with progressively smaller effective $\Delta({\bf k})$.

These differences are sufficient that a 
point contact or tunneling experiment performed along these lines should be able to distinguish them.
If there is no $\cos(k_z)$ dependence to the order parameter, the dI/dV curves
for these varying gate voltages should be essentially identical (up to a scale factor).

For simplicity, we have chosen above the real-space orientation orientation of the graphite such that the hexagonal face parallels the interface.  Substantially different dI/dV results obtain if instead the hexagonal face is perpendicular to the interface, as shown in the k-space Figure 2.  In this case the 30$^o$ rotation of the Brillouin zone relative to the real-space unit cell means that the graphite Fermi lines no longer occur at $k_{b}=0$, but are displaced above and below by approximately 0.737/$\AA$, which is relatively near the Sr$_{2}$RuO$_{4}$ $\gamma$ band k$_{F}$ of 0.75/$\AA$, so that the perfect order parameter sign change described above does not occur and one does not see a ZBCP.  Results for this case are presented in Figure 4, for the same parameters as in Fig. 3, and as in the previous plot substantial gate-voltage created differences are apparent, with the Andreev limit peak at $V=\Delta_{0}$ reducing with
increasing gate voltage, while in the tunneling limit substantial sub-gap density-of-states appear with increasing gate voltage.
\begin{figure}[h!]
\includegraphics[width=7cm]{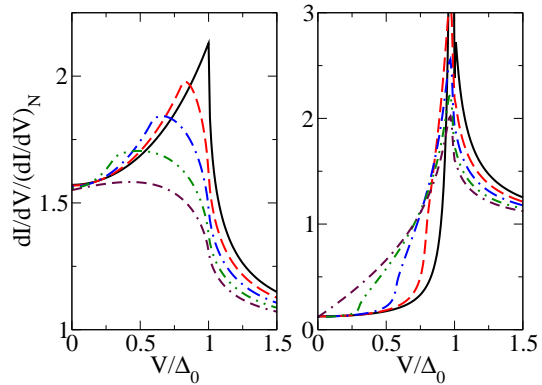}
\caption{The results of dI/dV calculations in the Andreev limit (left, Z=0.3) and tunneling limit (right, Z=2), for graphite gated and rotated 30$^o$ relative to that of Fig. 3 are shown.  Same parameters as in Fig. 3.}
\end{figure}

It is clear that these methods remain applicable even if the $k_{z}$ dependence of the 
order parameter is $a+b \cos(k_{z})$, with $|a| < b.$  The dI/dV curves will still evolve with gate voltage.  The main effect of such a constant term would be to change the gate voltage at which the Fermi surface cigar accesses the nodal lines, with a minor secondary effect on the shape of the DOS, so long as $a \ll b.$

{ \it Discussion - Experimental Consideration: Surface degradation.}
The results of the above section suggest that appropriate Andreev or tunneling experiments
may be able to determine whether or not there are nodes in the gap function located at $k_{z}=\pm \pi/2$, as an $\exp(i\phi)\cos(k_{z})$ order parameter would contain.There is, however, an experimental consideration requiring consideration: the surface degradation effects in this material.

There are to date three spectroscopic measurements made on 
Sr$_{2}$RuO$_{4}$ \cite{jin,upward}.  The first, 
by R. Jin et al \cite{jin}, performed tunneling on a Pb-Sr$_{2}$RuO$_{4}$ 
junction but found only a spectrum strongly resembling that of superconducting
lead, with coherence peaks (assumed to be that of Pb) at approximately 1.4 meV 
and no sub-gap structure.  The second experiment, by Upward et al \cite{upward}
performed c-axis STM using a Pt/Ir tip and found clear evidence of a superconducting gap.  The third experiment, by Laube et al \cite{lohneysen}, observed a weak Andreev reflection signal, as well as a zero-bias anomaly. 
Of issue
for our proposed experiment is the extremely large zero-bias
conductance - approximately 0.85 the normal-state value in the Upward et al data.  This is believed to result from surface degradation. 

A detailed accounting for the possible effects of surface degradation on tunneling or Andreev spectra is rather complicated, particularly as the cause of such degradation remains unknown.  Rather than attempt a first principles calculation as such, we therefore content ourselves with two relatively simple mechanisms for simulating the {\it effects} of surface degradation.

The first method is empirically based upon the comparative smallnessÊ (i.e. relative to background conductance) of the features observed in the spectroscopic work performed to date, and the lack of detail in these features.  Both of these characteristics are consistent with a large effective ``smearing" of the conductance.  To a lesser degree, such smearing is nearly universally present in spectroscopy on unconventional superconductors, where it is typically modeled by adding an imaginary part $\Gamma$ to the quasiparticle energy \cite{dynes}.   In fact such a $\Gamma$Ê has already been taken to represent the effects of a disordered surface layer \cite{chalsani}.  Given the predominance of this ``smearing" in experimental spectra of unconventional superconductors, and the comparative success of the Dynes $\Gamma$ in modeling such smearing, it is reasonable to use this parameter to simulate the as-yet-unknown source of surface disorder in Sr$_{2}$RuO$_{4}$.

I have performed a calculation, for the graphite orientation used in the main panels of Fig. 3,  with a large Dynes $\Gamma$ of 0.75$\Delta_{0}$; results of
this calculation are presented
in the insets of Fig. 3.   Even with this large $\Gamma$, the figures continue to show substantial 
differences in the dI/dV curves.  The zero-bias conductance changes significantly from $\delta\mu=-20meV$ to $-12.7 meV$ in both the Andreev and tunneling regime, as does the shape and size of the Andreev reflection signal, so that even if the problems
with surface degradation persist, the nodal lines, if existent, will leave
their distinctive signature in dI/dV.

A second method of simulating the effects of surface disorder is to relax the assumption of perfect parallel 
momentum conservation.  The premise here is that the surface degradation causes scattering within
the interface region, so that a finite, rather than infinitesimal region of superconductor Fermi surface is allowed to receive
current from a given graphite Fermi surface location.  In practice one may use a Gaussian distribution of conductance (i.e. $\propto (1+|a|^2-|b|^2)\exp(-\alpha(k_{\parallel}-k_{0,\parallel})^2$, where k$_{0,\parallel}$
\begin{figure}
\includegraphics[width=8cm]{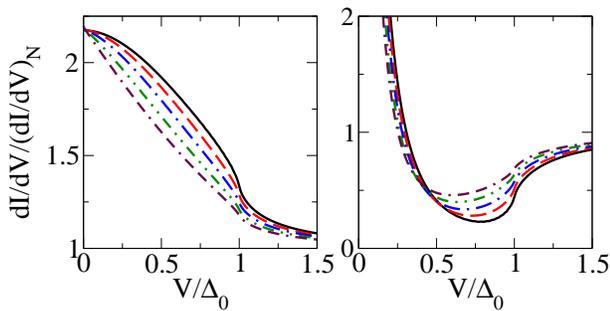}
\caption{dI/dV curves assuming significant nonconservation of parallel momentum, as described in the text.  Same parameters as figure 3, top (left here) and bottom (right).}
\end{figure}represents the parallel momentum of the incoming electron and k$_{\parallel}$ represents the 
parallel momentum of the transmitted quasiparticles,  and 
observe how the dI/dV features change.   We note that the boundary conditions of the Calculation section do not explicitly involve the parallel momentum, so this scheme is internally consistent.  The results are shown in Figure 5, 
where we show several Andreev limit dI/dV curves assuming an $\alpha$ of 
$1/\sqrt{0.2\pi}$ (i.e. $\Delta k_\parallel \sim 0.2\pi$) in units where $2\pi$ is $2\pi/c$, c the Sr$_2$RuO$_4$ lattice constant of 12.72 $\AA$.  Here $\delta k_{\parallel}$
is of the same order as typical $k_\parallel \sim \pi/2$, so that the non-conservation of parallel momentum is substantial.  We have also allowed for non-conservation of k$_{\parallel}$ in the k$_{b}$ direction, i.e. perpendicular to the graphite Fermi line, and taken account of the $\Delta({\bf k})$ phase factors that result. Even in this case, significant differences in dI/dV remain, so
that the proposed experiment can be considered robust against the surface degradation issue.

%One final issue, related to the problem of surface degradation, is that of the ab-plane orientation of the %graphite relative to Sr$_{2}$RuO$_{4}$.  The most natural orientation for tunneling into a material as %two dimensional as graphite is along the c-axis, while that proposed here is into the ab-plane.  At first %sight this  experimental proposal may therefore seem naive or impractical.  Recently, however, Gonelli %and co-workers \cite{gonelli}  have demonstrated an innovative technique for forming point contacts %on the ab-plane of the graphite-intercalated superconductor CaC$_{6}$ by use of a small drop of Ag %conductive paint.  Clear evidence of a superconducting gap was observed in this experiment, as well as in similar
%experiments \cite{gonelli2} performed on MgB$_{2}$, thus demonstrating the quality of such contacts.
%An experiment forming an ab-plane oriented point contact between graphite and Sr$_{2}$RuO$_{4}$ %should therefore be feasible.

{\it Conclusion.  }
In this work I have demonstrated that a series of superconductive NIS tunneling
measurements on Sr$_{2}$RuO$_{4}$ using gated graphite as the normal
electrode should allow determination of the presence or absence of order parameter
line nodes parallel to the ab-plane.  I have shown that 
the power of such an experiment to determine nodal structure is robust against the surface disorder prevalent in this material.  I await the results of such experiments with
great interest. 

{\bf Acknowledgements} It is a pleasure to acknowledge a valuable discussion 
with D.J. Van Harlingen and helpful interactions with Y. Tanaka and N. B. Ali.  In addition, I wish to thank M. D. Johannes and I.I. Mazin for their reading of the manuscript prior to submission.

\end{document}